\documentclass[aip,
jmp,
 amsmath,amssymb,
 reprint,%
]{revtex4-1}

\usepackage{graphicx}% Include figure files
\usepackage{dcolumn}% Align table columns on decimal point
\usepackage{bm}% bold math
\usepackage[utf8]{inputenc}
\usepackage[T1]{fontenc}
\usepackage{mathptmx}
\usepackage{etoolbox}
\usepackage[colorlinks=true]{hyperref}

\makeatletter
\def\@email#1#2{%
 \endgroup
 \patchcmd{\titleblock@produce}
  {\frontmatter@RRAPformat}
  {\frontmatter@RRAPformat{\produce@RRAP{*#1\href{mailto:#2}{#2}}}\frontmatter@RRAPformat}
  {}{}
}%
\makeatother

\begin{document}
% -------------------------------------------------------------------- %

% the following line is for submission, including submission to the arXiv!!
%\hspace{5.2in} \mbox{}

\preprint{AIP/123-QED}

% ----- title and authors -------------------------------------------- %

% -- title
\title[Sub-tesla on-chip nanomagnetic metamaterial platform for angle-resolved photoemission spectroscopy]{Sub-tesla on-chip nanomagnetic metamaterial platform for angle-resolved photoemission spectroscopy}

% -- authors
\author{Wenxin Li}
    \affiliation{Department of Applied Physics, Yale University, New Haven, Connecticut 06511, USA}
\author{Wisha Wanichwecharungruang}
    \affiliation{Department of Physics and Astronomy, Rice University, Houston, TX, 77005, USA}
\author{Mingyang Guo}
    \affiliation{Department of Physics, Boston College, Chestnut Hill, Massachusetts 02467, USA}
\author{Ioan-Augustin Chioar}
    \affiliation{Department of Applied Physics, Yale University, New Haven, Connecticut 06511, USA}
    \affiliation{Department of Physics, Princeton University, Princeton, NJ 08540, USA}
    \affiliation{Department of Physics and Astronomy, University of Maine, Orono, ME 04469, USA}
\author{Nileena Nandakumaran}
    \affiliation{Department of Chemical Engineering and Materials Science, University of Minnesota, Minneapolis, Minnesota 55455, USA}
\author{Justin Ramberger}
    \affiliation{Department of Chemical Engineering and Materials Science, University of Minnesota, Minneapolis, Minnesota 55455, USA}
\author{Senlei Li}
    \affiliation{School of Physics, Georgia Institute of Technology, Atlanta, Georgia 30332, USA}
\author{Zhibo Kang}
    \affiliation{Department of Applied Physics, Yale University, New Haven, Connecticut 06511, USA}
\author{Jinming Yang}
    \affiliation{Department of Physics, Yale University, New Haven, Connecticut 06511, USA}
\author{Donghui Lu}
    \affiliation{Stanford Synchrotron Radiation Lightsource, SLAC National Accelerator Laboratory, Menlo Park, California 94025, USA}
\author{Makoto Hashimoto}
    \affiliation{Stanford Synchrotron Radiation Lightsource, SLAC National Accelerator Laboratory, Menlo Park, California 94025, USA}
\author{Chunhui Rita Du}
    \affiliation{School of Physics, Georgia Institute of Technology, Atlanta, Georgia 30332, USA}
\author{Chris Leighton}
    \affiliation{Department of Chemical Engineering and Materials Science, University of Minnesota, Minneapolis, Minnesota 55455, USA}
\author{Peter Schiffer}
    \affiliation{Department of Applied Physics, Yale University, New Haven, Connecticut 06511, USA}
    \affiliation{Department of Physics, Princeton University, Princeton, NJ 08540, USA}
\author{Qiong Ma}
    \affiliation{Department of Physics, Boston College, Chestnut Hill, Massachusetts 02467, USA}
\author{Ming Yi}
    \affiliation{Department of Physics and Astronomy, Rice University, Houston, TX, 77005, USA}
    \affiliation{Smalley-Curl Institute, Rice University, Houston, TX, 77005, USA}
\author{Yu He}
    \email{yu.he@yale.edu}
    \affiliation{Department of Applied Physics, Yale University, New Haven, Connecticut 06511, USA}

% -- date
\date{\today}

% ----- 0 abstract --------------------------------------------------- %
\begin{abstract}
Magnetically controlled states in quantum materials are central to their unique electronic and magnetic properties. However, direct momentum-resolved visualization of these states via angle-resolved photoemission spectroscopy (ARPES) has been hindered by the disruptive effect of magnetic fields on photoelectron trajectories. Here, we introduce an \textit{in-situ} method that is, in principle, capable of applying magnetic fields up to 1 T. This method uses substrates composed of nanomagnetic metamaterial arrays with alternating polarity. Such substrates can generate strong, homogeneous, and spatially confined fields applicable to samples with thicknesses up to the micron scale, enabling ARPES measurements under magnetic fields with minimal photoelectron trajectory distortion. We demonstrate this minimal distortion with ARPES data taken on monolayer graphene. Our method paves the way for probing magnetic field-dependent electronic structures and studying field-tunable quantum phases with state-of-the-art energy-momentum resolutions.
\end{abstract}

\maketitle

% ----- 1 introduction ----------------------------------------------- %
\section{Introduction}

% --- FIGURE 1 --- %
\begin{figure}
\includegraphics[width=1\columnwidth,clip,angle =0]{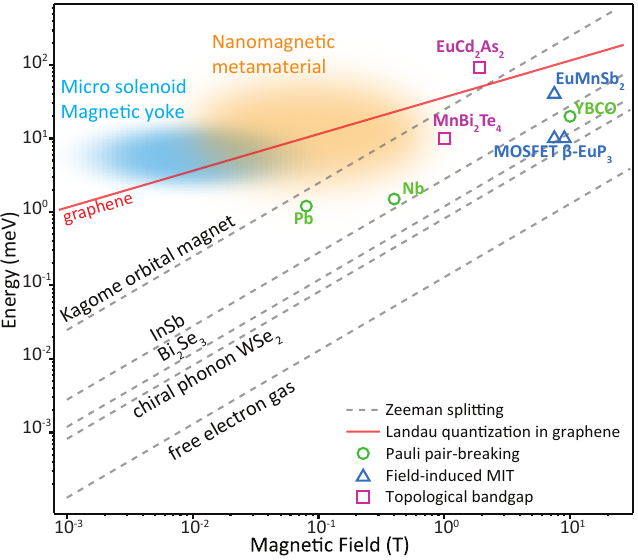}
\caption{\label{fig1}
    Examples of solid-state phenomena under magnetic field (see main text for details) and their corresponding energy scale. The blue and orange shaded areas indicate the accessible field strength and energy resolution of photoemission under magnetic field in previous approaches and the method proposed in this work, respectively.
}
\end{figure}

% [a] 
Manipulating the phases of matter has been a central theme in condensed matter and materials physics. Among the various external stimuli, magnetic fields have emerged as a particularly powerful means to tune material properties, including inducing quantum phase transitions \cite{Rev_magphasetran}, altering superconducting states \cite{aoki2019review}, as well as driving spin reorientation phenomena \cite{Rev_Spintronics}. Magnetic fields also provide powerful control over topological states of matter, thanks to their ability to drive closed-loop electron motion in both real and momentum space~\cite{Rev_TISC,xiao2010berry}. However, detecting magnetic-field-mediated electronic behavior directly in energy-momentum space remains extremely challenging, in part due to the minuscule energy scales associated with typical magnetic fields realized in a laboratory setting.
Figure~\ref{fig1} compiles the energy scales of various emerging solid state phenomena under an external magnetic field \cite{BiSe,Chiralphonon,Deng2021MBT,EuCdAs,InSb,SC-1,SC-2,SC-3,Sun2021EuMnSb,Wang_2020EuP}. The gray dashed lines represent magnetic field-induced Zeeman splitting, an effect linearly proportional to the applied magnetic field; the splitting energy $E = g{\mu_B}B$, where $g$ is the Land\'e $g$ factor and $\mu_B$ is the Bohr magneton. While $g$ = 2 is the most common case in electron-spin-dominated systems, emergent orbital magnetism can enhance $g$ by two orders of magnitude \cite{BiSe,Chiralphonon,InSb}. The green circles represent examples of the pairing energies and thermodynamic critical fields in superconductors, which follow the relation $\frac{1}{2}\rho(E_F) \Delta^2 \sim B^2/(2\mu_0)$~\cite{SC-1,SC-2,SC-3}, where $\rho(E_F)$ is the electronic density of states at the Fermi level, and $\Delta$ is the superconducting energy gap. The blue triangles show examples of magnetic field-induced metal-to-insulator transitions~\cite{Sun2021EuMnSb,Wang_2020EuP} and the purple rectangles show examples of gap modulation of topological materials \cite{Deng2021MBT,EuCdAs}. The red solid line shows the Landau quantization of a Dirac fermion, which is the key to understanding many of its unique electronic and magnetic properties~\cite{GRLLreview,GRgeneral1,GRgeneral2,GRgeneral3,GRgeneral4}.
%\\

% [b] 
To truly understand and engineer the tunability of many magnetically active materials, it is desirable to investigate their electronic structure in the presence of an external magnetic or electric field. Traditionally, this is done with high-field electrical transport, where quantum oscillations are used to derive the underlying Fermi surface geometry~\cite{QOFermi-1,QOFermi-2}. In the recent two decades, angle-resolved photoemission spectroscopy (ARPES) has become a mainstream method to probe the electronic structure of quantum materials \cite{ARPESreview2003,ARPESreview2021}.
However, this technique is notoriously vulnerable to external electromagnetic fields. Recent developments in ARPES have demonstrated \textit{in-situ} active electrostatic gating or voltage biasing in the sample environment, to either investigate gating effects in heterostructures \cite{Gate-ARPES-2019,Gate-ARPES-2020,Gate-ARPES-2021} or effectively expand the momentum field of view \cite{BiasARPES}.
In contrast to the rapid progress in incorporating static electric fields, applying an \textit{in-situ} magnetic field to samples undergoing ARPES measurements remains a major challenge because the field deflects photoelectrons, leading to unwanted effects, including emission angle contraction, constant energy contour (CEC) rotation, and momentum broadening of the ARPES spectrum~\cite{B-ARPES-Rice}.
Moreover, magnetic fields are much harder to confine than electrostatic fields because of their divergence-free nature. As such, most ARPES systems are designed to shield the magnetic field to milligauss levels so that electrons with eV-scale kinetic energy have deflections below the typical angular resolution ($\leq0.1^\circ$).
%\\

Recently, exciting efforts have been made to apply an \textit{in-situ} magnetic field for ARPES measurements, using either a magnetic yoke device \cite{B-ARPES-ALS} or a solenoid at the sample position \cite{B-ARPES-Rice}. Notwithstanding the demonstration of feasibility, photoelectron trajectories are still strongly affected by the seemingly unavoidable far-field magnetic field.
With the solenoid design, a 3~mT out-of-plane field already causes 5\% emission angle contraction and $22^\circ$ CEC rotation \cite{B-ARPES-Rice}.
With the yoke device design, the direction of the generated magnetic field varies rapidly in space~\cite{B-ARPES-ALS}. This spatial inhomogeneity limits the utility of the setup to carefully designed nano-ARPES systems capable of sub-micron level position reproducibility.
These detrimental far-field deflection effects limit the acceptable field strength in ARPES measurements to well below 100~mT, making it extraordinarily challenging to induce detectable physical effects above the state-of-the-art ARPES energy resolution threshold in typical quantum materials (Fig.~\ref{fig1}).
This calls for new approaches to realize an \textit{in-situ} magnetic field sample environment.
%\\

In recent years, nanoscale metamaterials have dramatically reshaped the landscape of quantum materials research. By enabling unprecedented control over electromagnetic environments at the nanoscale, they have opened new avenues for manipulating quantum phases and probing emergent phenomena. A significant body of work has focused on nanophotonic substrates to achieve high-frequency electric field confinement, leading to discoveries such as electron–hole liquid formation~\cite{Photon-MoS2}, spatially modulated excitonic textures~\cite{Photon-Youzhou}, and magnetically tunable strong light–matter coupling in van der Waals antiferromagnets~\cite{Photon-comin2025}.
However, the magnetic counterpart—nanoscale magnetic field engineering—remains comparatively underexplored.
In this work, we introduce a nanomagnetic metamaterial platform for applying an \textit{in-situ} magnetic field in ARPES measurements: using nanomagnet arrays with alternating polarity. This method should allow for $\sim10$ mT to $\sim1$ T near-fields and a diminishing far-field, keeping the photoelectron deflection below $0.1^\circ$. We
also perform proof-of-concept ARPES measurements with such an \textit{in-situ} magnetic field on monolayer graphene, and provide a general design guideline for future implementation of such nanomagnetic sample environments.

% ----- 2 results and discussion ------------------------------------- %
\section{Results and discussion}

% ----- simulation part ------------------------------------- %
\subsection{Device design and simulation of effects on ARPES spectrum}

% --- FIGURE 2  --- %
\begin{figure}
\includegraphics[width=1\columnwidth,clip,angle =0]{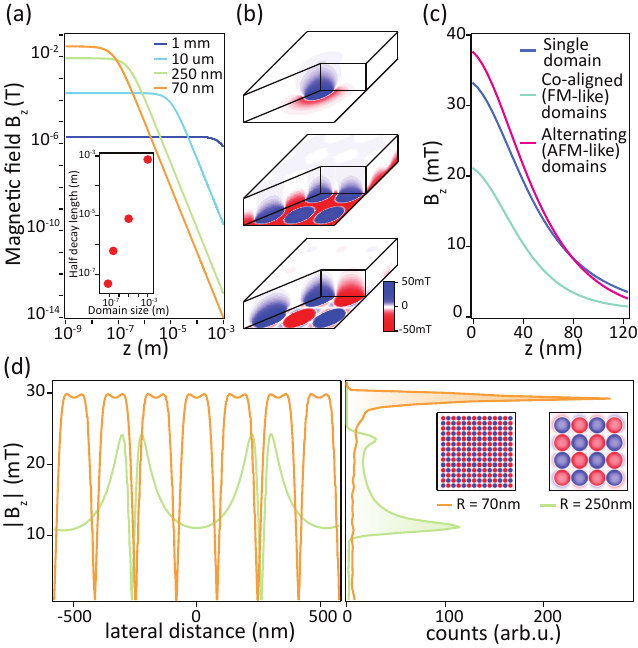}
\caption{\label{fig2}
    Calculated magnetic fields from different magnet configurations. (a) $z$-axis magnetic field ($B_z$) generated by cylindrical magnets with radii $R =$ 1 mm, 10 $\mu$m, 250 nm, and 70 nm, evaluated along the central axis of the cylinder. $z$~=~0 represents the top surface of the magnet. (b) Intensity plot of magnetic field distribution above one single-domain magnetic island (top), a nanomagnet array with all islands magnetized in the same direction (middle) and a nanomagnet array with alternating polarity (bottom) respectively. $R=70$~nm and an edge-to-edge gap of 25~nm are used. (c) Comparison of $B_z$ with respect to the distance from the top plane of the magnet between different scenarios shown in (b). (d) Comparison of $|B_z|$ distribution and histogram at 30~nm above the nanomagnet array between $R=70$~nm and $R=250$~nm.
}
\end{figure}

% --- FIGURE 3 (photoelectron trajactory) --- %
\begin{figure*}
    \includegraphics[width=1.95 \columnwidth,clip,angle =0]{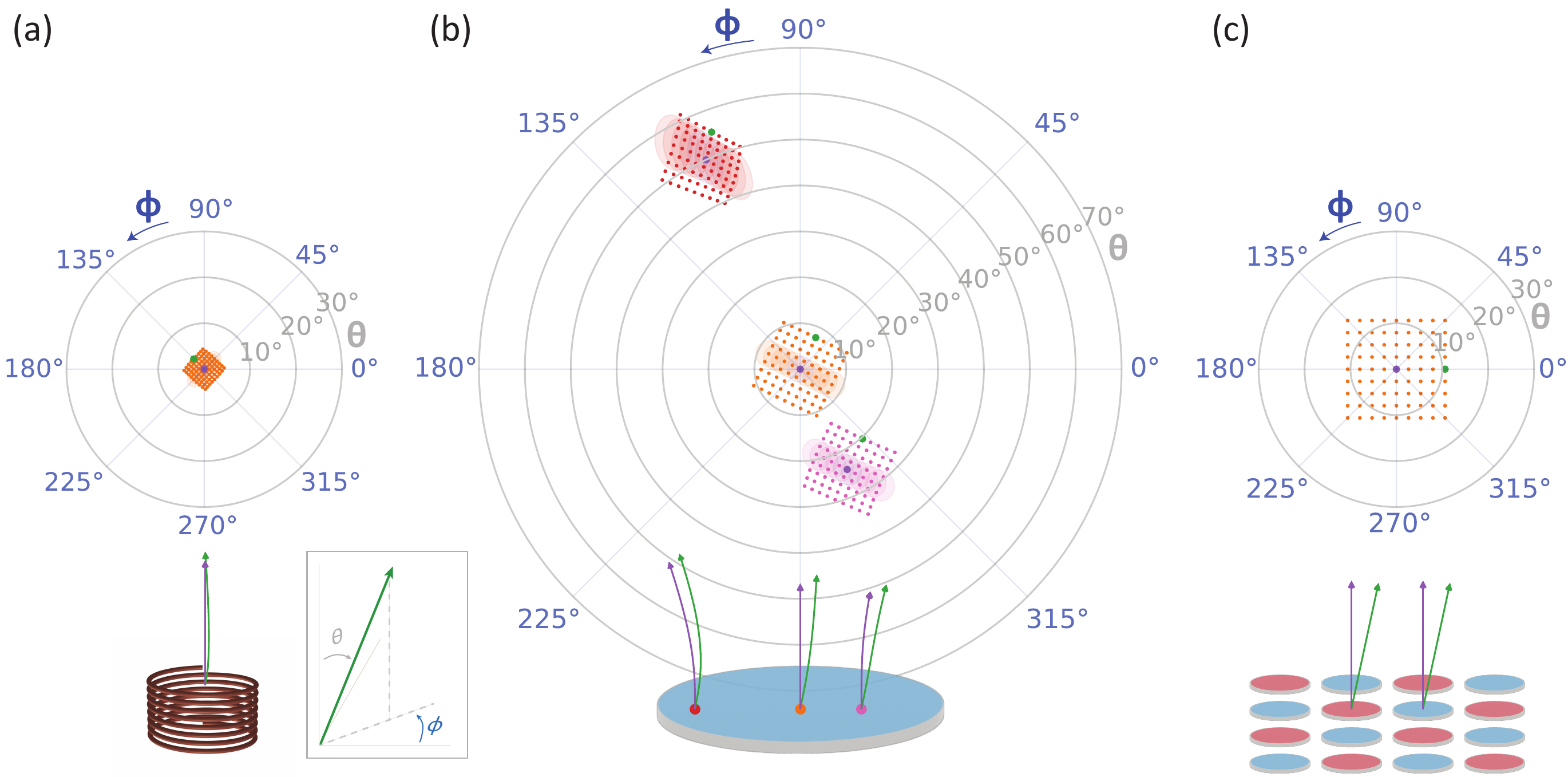}
    \caption{\label{fig3} 
    Simulated photoelectron trajectory and emission angle distribution maps under different magnetic fields. (a) A solenoid setup used in \cite{B-ARPES-Rice}. 
    (b) A single cylindrical magnet with radius $R = 1$ mm, showing representative positions at distance $d = 0$ (orange), 0.4 mm (purple), and 0.7 mm (red) from the center. Semi-transparent circles illustrate the broadening effect (see SI~\cite{SI}).
    (c) A nanomagnet array with alternating polarity, as proposed in this work.
    In all cases, the out-of-plane magnetic field is set to $|B_z|=30$~mT at $z$~=~30~nm. Grids in all panels represent the photoelectron emission angle distribution maps. Violet and green arrows correspond to the trajectories of the marked dots in the grids (see the main text for details).
}
\end{figure*}

% [a] 
The key innovation of this setup compared to existing setups \cite{B-ARPES-ALS,B-ARPES-Rice} is the use of a periodic array of magnetic dipoles (multipole magnets) instead of a single dipolar magnet.
For a macroscopic single-domain magnet, the self-depolarization effect would render a rapidly decaying field distribution from the edge to the center of the magnet. Moreover, such a single-domain magnet would effectively behave like a current loop, yielding extensive far-fields along all directions at the scale of its lateral dimensions.
On the other hand, with an array of alternating poles, the self-depolarizing effect is reduced to enhance near-field strength, and the far field is rapidly canceled to $|\Vec{r}-\Vec{r'}|^{-l}$, where $|\Vec{r}-\Vec{r'}|$ is the distance to the observation point and $l$ is the order of the multipole~\cite{Jackson}. This is most evident through the Halbach arrays \cite{HALBACH1980,halbach1985} used in fridge magnets to create a strong, highly confined, single-sided magnetic field. 
%\\

To quantitatively demonstrate the advantage of using multipole magnets as a magnetic field source and a substrate to the sample, and to find the best approach to provide a strong and highly confined magnetic field, we perform magnetostatics calculations. We assume uniform magnetization for each magnetic island in all following discussions. We first calculate the magnetic field of a single island as a function of its radius. 
The magnetic field emanating from a finite-size permanent magnet can be calculated using the scalar potential~\cite{Jackson}. The magnetic scalar potential $\Phi$ at position $\vec{r}$ for a uniformly magnetized object is:
\begin{equation}
% \Phi(\Vec{r}) = \frac{1}{4\pi} \oint_{V'} \frac{-\nabla \cdot \Vec{M}}{|\Vec{r}-\Vec{r'}|} dV'
\Phi(\vec{r}) = \frac{1}{4\pi} \int_V \frac{-\nabla' \cdot \vec{M}(\vec{r}')}{|\vec{r} - \vec{r}'|} \textrm{d}^3 \vec{r}'
\end{equation}
where $M$ is the magnetization of the material and the integration is over the volume of the material.
For a cylindrical magnet with radius $R$ (in the $x$-$y$ plane) and height $h$ (along the $z$ axis), the $B_z$ distribution along the central axis of the magnet is given by:
    \begin{equation} \label{eq2}
        B_z(z)=\frac{\mu_0M_z}{2}\left[\frac{z+h}{\sqrt{R^2+\left(z+h\right)^2}}-\frac{z}{\sqrt{R^2+z^2}}\right]
    \end{equation}
Here, the magnet spans from $z=-h$ to $z=0$, with $z=0$ denoting the top surface. 
%
%\\

\begin{figure}
    \includegraphics[width=1\columnwidth,clip,angle =0]{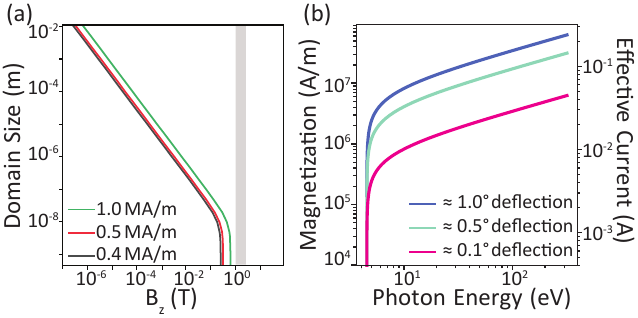}
    \caption{\label{fig5} 
    Design guidelines. (a) Island size required to achieve the desired $B_z$ field for different material magnetization strengths. (b) Maximum magnetization (left axis) and the corresponding effective current (right axis) for which the photoelectron deflection stays below certain thresholds---note that the total deflection does not depend on the island size (see SI~\cite{SI}).
}
\end{figure}

Figure \ref{fig2}~(a) compares the $B_z$ distribution along the central axis of cylindrical magnets with different radii $R$. The calculation is performed with the top surface of the magnet centered at (0,0,0), the height of the magnet set to $h$~=~10~nm, and the magnetization of the magnet set to $M_z$~=~0.32~MA/m, a value close to real-world materials, as will be discussed in Fig.~\ref{fig5}.
Clearly, the smaller the island size, the larger the near-field and the smaller the far-field magnetic field it will generate.
%
%\\

However, the magnetic field generated by a single domain decays at the length scale of its lateral dimensions, making it less suitable for photoemission experiments, where the sample size would typically fall in the micron to millimeter range.
Thus, in this setup, an array consisting of multiple magnetic islands is necessary. Unlike typical Halbach arrays with an in-plane polarization~\cite{halbach1985,HALBACH1980}, in order to achieve an out-of-plane field, the polarity of each island needs to point either into or out of the sample plane (i.e., along the $z$ axis). To find the optimal array arrangement, we compare the $B_z$ distribution in three different scenarios: one single-domain island in free space, multiple islands magnetized in the same direction, and multiple islands with an alternating polarity, as shown in Fig.~\ref{fig2}~(b) and (c).
To compute the off-axis magnetic field and aggregate contributions from multiple magnetic islands, we approximate each island as a current loop—valid under the condition that the island height is much smaller than its radius, which applies to the scenarios discussed below. In this approximation, $B_z$ in cylindrical coordinates is given by:
\begin{equation} \label{eq3}
\begin{split}
    B_z &=\left(\frac{R^2 - s^2 - z^2}{(R-s)^2 + z^2}E_E(k) + E_K(k)\right) f 
\end{split}
\end{equation}
where $f \equiv \frac{\mu_0 I}{2 \pi} \frac{1}{\sqrt{(R+s)^2+z^2}}$, $k \equiv \frac{4 R s}{(R+s)^2 + z^2}$, $I=Mh$ is the effective current of the magnetic island, $z$ and $s$ are the vertical and radial distances from the loop center to the point of interest, and $E_K$ and $E_E$ are the complete elliptic integrals of the first and second kinds, respectively.
The island size in all three scenarios is set to be $R$~=~70~nm, which is practically achievable when fabricating nanomagnet arrays~\cite{CoPyIslandSize,CoCrPtIslandSize,CoPt-Schiffer2012}.
An array consisting of 14~$\times$~14 islands is used for simulating the latter two scenarios.
The gap between the nearest-neighbor islands is set to be 25 nm, which is realistic with modern-day fabrication technologies~\cite{fabgap-1,fabgap-2,fabgap-3}. Figure~\ref{fig2}~(c) shows that an array of islands with alternating polarity produces the strongest near-field $|B_z|$, which remains almost constant away from the surface up to its linear dimension, and then rapidly decays beyond this point. This makes this setup particularly applicable to bulk samples with thicknesses under a micron.
%\\

In addition to the field strength, spatial homogeneity of the field is also desired. Figure~\ref{fig2} (d) shows the lateral distribution (left panel) and histogram (right panel) of the field strength $|B_z|$ at 30 nm above the nanomagnet array, comparing a 14~$\times$~14 array with island size $R$ = 70 nm, and a 4~$\times$~4 array with $R$ = 250 nm. A smaller island size is found to yield not only a stronger overall near‑field, but also a more homogeneous magnetic‑field magnitude across the $x$-$y$ plane when averaged over the typical multi‑micron ARPES beam spot.
%\\

% [a]
Having achieved a uniform, large $|B_z|$ at the sample, we turn to investigate deflection effects, i.e., how this magnetic field will affect the photoelectron trajectory. We perform simulations of the photoelectron emission angle distribution maps under three different sample environment designs -- a macroscopic solenoid~\cite{B-ARPES-Rice}, a single uniformly magnetized island, and a nanomagnet array with alternating polarity (Fig.~\ref{fig3}, see Supplementary Information (SI) for detailed methods~\cite{SI}). An electron kinetic energy of 16.9 eV is used, which typically corresponds to photoelectrons ejected with 21.2 eV photons (He I$\alpha$ line from a plasma UV lamp). An evenly spaced photoelectron emission angle distribution map of $10.6^\circ\times10.6^\circ$ (i.e., $\frac{15^\circ}{\sqrt{2}}$) is created to represent the emission angle distribution without a magnetic field. The radial emission angle is defined as $\theta$ and the in-plane azimuthal angle is defined as $\phi$. Photoelectrons with emission angle $(\theta,\phi)=(0,0)$ and ($10.6^\circ$, 0) are marked with violet and green dots respectively. To ensure comparability, the near-field (field at $z$~=~30~nm above the nanomagnet array) is maintained at $|B_z|=$ 30 mT in all setups, and the magnetic field along all three directions ($B_x$, $B_y$, $B_z$) is considered in the analysis. 
%\\

The solenoid method causes an obvious CEC in-plane rotation of almost $135^\circ$ and an overall emission angle contraction of more than 50\%, as shown from the positions of the violet and green dot in Fig.~\ref{fig3}~(a). For a large single dipolar magnetic film substrate [Fig.~\ref{fig3}~(b)], the magnetic field varies significantly from the edge of the substrate to the center, with the strongest field observed at the edge. This variation is evident from the pronounced and spatially inhomogeneous distortion in the trajectories of photoelectrons emitted from the center of the magnet (orange), at $d = 0.4$ mm from the center (purple), and at $d = 0.7$ mm from the center (red). This makes the method shown in (b) highly challenging in real applications. For the alternating nanomagnet array [Fig.~\ref{fig3}~(c)], the residual magnetic field at 50 $\mu$m away from the sample is well below $10^{-7}$ T. The associated CEC rotation effect is only $0.04^\circ$, and the angular contraction is less than 0.2\%, well below the state-of-the-art angular resolution of photoelectron analyzers.

In all, building upon the magnetic field calculations, we demonstrate the feasibility of using a nanomagnet array with alternating polarity to move the accessible field intensity one to two orders of magnitude larger (blue to orange shades in Fig.~\ref{fig1}) while maintaining nearly negligible size and spatial inhomogeneity of photoelectron deflection.
An important caveat of our approach is that the field direction alternates between up and down across the magnetic islands, limiting this setup to the study of magnetic phenomena that are symmetric with respect to $\pm B_z$---for example, field‑induced phase transitions and Landau‑level splitting in most inversion‑symmetric or nonmagnetic materials.

% --- FIGURE 5 --- %

% --- FIGURE 6 --- %
\begin{figure*}[!t]
    \includegraphics[width=2\columnwidth,clip,angle =0]{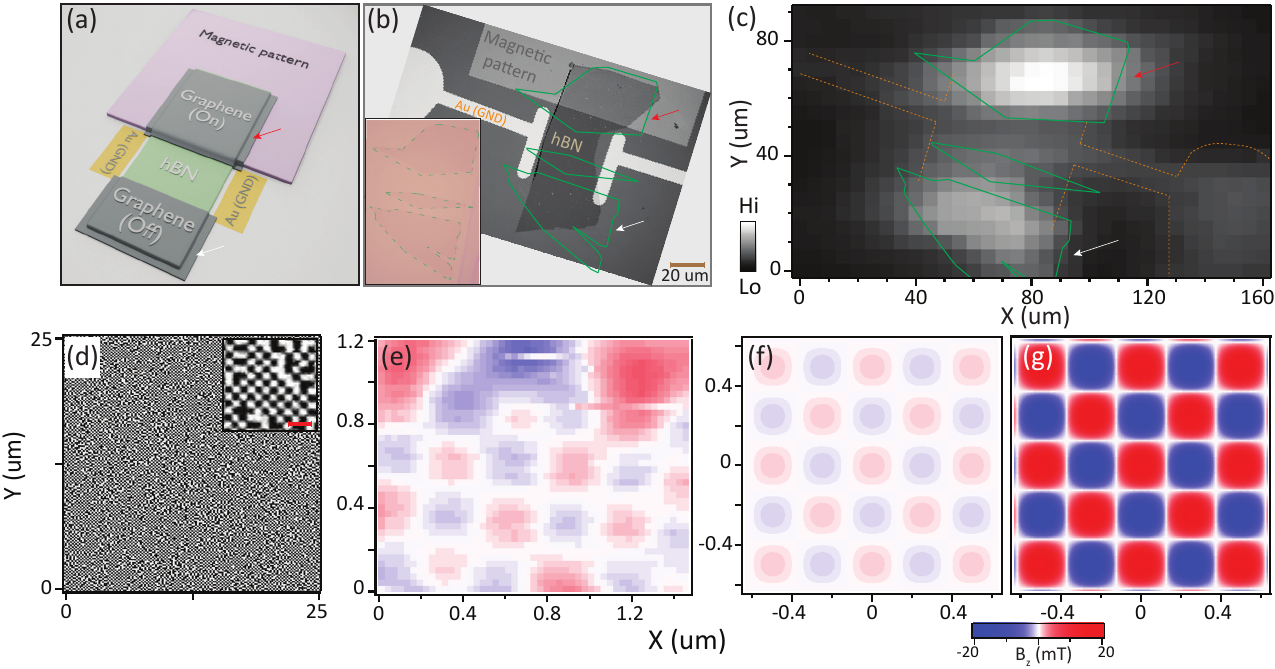}
    \caption{\label{fig6} 
    Experimental setup and stray field characterization. (a)~Device schematic. (b)~Optical image of the device. Green outlines indicate graphene boundaries on (red arrow) and off (white arrow) the magnetic pattern (light gray, top). The darker region corresponds to the hBN, and the white area to the gold contact pad. The inset (lower left) shows a magnified optical image of the graphene. (c)~ARPES real-space scan overlaid with graphene boundaries, showing clear signals on/off the magnetic pattern and on the gold pad. (d)~MFM image of the array of $R$ = 70 nm nanomagnets. Each island in the MFM image appears either black or white, indicating magnetic moments pointing upward and downward respectively. The inset shows a zoomed-in view of the MFM image. The scale bar (red) in the inset represents 0.5 $\mu$m. (e) $B_z$ map from scanning NV magnetometry. (f) Simulated $B_z$ map at 200~nm above the sample. (g) Simulated $B_z$ map at the graphene layer.
}
\end{figure*}

To enable applications across diverse material systems, we outline a general design principle for nanomagnet arrays tailored to different experimental needs.
Such arrays can be realized by leveraging advances in artificial spin ice (ASI)—metamaterials composed of coupled nanomagnets arranged on engineered lattices~\cite{ASIreview1,ASIreview2,ASIreview3}.
This metamaterial approach builds on well-established protocols and widely accessible nanofabrication techniques, with previous ASI studies demonstrating the successful formation of ordered islands with alternating polarity through rapid demagnetization protocols~\cite{CoPt-Schiffer2012,ASI-demag-1,ASI-demag-2}.
The key design parameter is the required out-of-plane magnetic field $B_z$ at the sample position, which is set by the magnetic material’s saturation magnetization $M_z$ and the array geometry.
Materials combining large perpendicular magnetic anisotropy (PMA) with high saturation magnetization are preferred for achieving stronger $B_z$~\cite{PMAreview}.

Figure~\ref{fig5}~(a) shows the relation between magnetic island size and the target $B_z$ for several representative PMA materials.
To eliminate the complexity introduced by varying edge-to-edge distances in different designs, the data presented in Fig.~\ref{fig5} are calculated using single-island configurations. As such, they represent a lower bound—yet remain within the same order of magnitude—for the stray field generated, as illustrated in Fig.~\ref{fig2}~(c).
Once the required field strength is specified, the substrate material and island size can be chosen directly from this plot.
In the interatomic limit, the configuration reduces to a purely classical analogue of an antiferromagnet.
The solid lines correspond to the design geometry used in the proof-of-concept device with a magnet thickness of $h = 10$~nm (see \hyperref[sec:experiment]{Section~B}).
For example, to realize $B = 0.1$~T, one could use Co/Pt multilayers ($M_s \sim 0.4$~MA/m~\cite{CoPt}) with an island size of $R=24$~nm, Tb/Co multilayers ($M_s \sim 0.5$~MA/m~\cite{TbCoMs,TbCoDose}) with $R=31$~nm, or FePt ($L1_0$ phase, $M_s \sim 1.0$~MA/m~\cite{PMA_FePt,FePtMs}) with $R=64$~nm.
This scaling relation provides a straightforward design rule, enabling flexible tuning of $B_z$ across different materials and fabrication constraints.

The gray-shaded region indicates the expected upper limit of $B_z$ attainable with this method. By optimizing the magnet geometry,
a maximum field of $\mu_0 M = 1.26$~T (lower bound of the gray-shaded region) can be reached with FePt in the $L1_0$ phase~\cite{PMA_FePt,FePtMs}.
Continued advances in interface engineering of PMA materials with higher saturation magnetization could push this further, with fields approaching $\mu_0 M = 2.45$~T (upper bound of the gray-shaded region). This value corresponds to the saturation magnetization of an Fe$_{0.65}$Co$_{0.35}$ alloy~\cite{Alloymagnetbook,FeCostrongestmagnet}.
Since both Fe/Pt and Co/Pt exhibit strong PMA~\cite{CoPt,PMA_FePt}, Fe$_{0.65}$Co$_{0.35}$/Pt may represent a promising candidate for high-field PMA systems—provided the anisotropy can be preserved in alloys with maximized saturation magnetization, despite the current lack of experimental reports.

%\\

Subsequently, the minimum photon energy can be determined on the basis of acceptable photoelectron angular distortions. 
Since the CEC rotation angle depends only on the magnetic material’s magnetization $M$ and the photoelectron kinetic energy (see SI~\cite{SI}), one can map combinations of $M$ and photon energy after setting a tolerance for photoelectron deflection.
Figure~\ref{fig5}~(b) shows this dependence for several CEC rotation thresholds, with the work function fixed at $\Phi = 4.5$~eV~\cite{ARPESreview2003,ARPESreview2021}.
When the choice of photon energy is crucial—for example, to optimize energy resolution or probe $k_z$ dispersion~\cite{ARPESkz1,ARPESkz2,ARPESkz3}—the required $M$ (or an equivalent effective current for loop geometries) can be chosen directly from Fig.~\ref{fig5}~(b).
In cases without strict photon energy requirements, material selection may instead be guided by fabrication compatibility, with the acceptable photon energy range determined from the same figure.
%\\
%

% ----- experiment part ------------------------------------- %
\subsection{Experimental demonstration} \label{sec:experiment}

Next, we present ARPES measurements conducted using this setup, with the objective of quantitatively assessing the photoelectron deflection effect inherent to the method. Monolayer graphene is selected as the benchmark system due to its extensively characterized electronic structure, featuring sharply defined and highly dispersive linear bands at the corners of the Brillouin zone.
Figure~\ref{fig6}~(a) shows a schematic of the device, with the corresponding optical image in Fig.~\ref{fig6}~(b).
The nanomagnetic metamaterial array consists of [Co(3\AA)/Pt(10\AA)]$_{12}$ multilayers, patterned into islands with radius $R = 70$ nm and edge-to-edge spacing of 25 nm~\cite{CoPt,CoPt-Schiffer2012}, on silicon wafers (details in SI~\cite{SI}).
Two graphene pieces, exfoliated from a single flake (green outlines in Fig.~\ref{fig6}~(b)), are placed on magnetic (red arrow) and non-magnetic (white arrow) regions, respectively, and are mechanically supported by an hBN flake (see SI for fabrication methods~\cite{SI}). Both graphene flakes are grounded via gold pads evaporated on the wafer, making the device compatible with ARPES experiments. Further ARPES experimental details are provided in SI~\cite{SI}.

An ARPES real-space map of the integrated photoemission intensity [Fig.~\ref{fig6}~(c)] resolves the shape of the two graphene flakes. The map agrees well with the optical image, enabling direct comparison of ARPES spectra taken on and off the nanomagnet array.
To verify the magnetic field distribution, we perform Magnetic Force Microscopy (MFM) and scanning Nitrogen-Vacancy (NV) magnetometry measurements~\cite{NV-GT2024,NV-review2014,NV-GT2023}.
Figure~\ref{fig6}~(d) shows an MFM image of the patterned magnetic array over a $25\mu\text{m}\times25\mu\text{m}$ field of view, revealing extended antiferromagnetically ordered domains. The inset presents a $2\mu\text{m}\times2\mu\text{m}$ zoomed-in region for higher-resolution visualization of the same ordering. The uniform contrast confirms effective demagnetization across the entire area and validates the choice of a compact square‑array design for effectively achieving alternating magnetic domain structures via nanomagnetic metamaterials.

To further verify the robustness of the magnetic structure following sample treatments (see SI for detailed procedures~\cite{SI}) and UV light exposure, and to quantify the magnetic field distribution,
Fig.~\ref{fig6}~(e) shows the stray field pattern measured by scanning NV magnetometry~\cite{NV-GT2024,NV-review2014} on the magnetic region of the graphene/hBN device after the ARPES experiments. The NV–sample distance was kept at 200~nm. Alternating magnetic domain structures are clearly visible, with an out-of-plane stray field magnitude $|B_z| \approx 1.3$~mT. Further NV magnetometry experimental details and data processing methods are provided in SI~\cite{SI}.
Figures~\ref{fig6}~(f) and (g) present simulated stray field patterns under the same geometry, using $M_s=300$~kA/m. The simulation yields $|B_z| = 1.02$~mT at 200~nm above the sample—consistent with the NV‑magnetometry measurement—and predicts an out‑of‑plane field of $|B_z| = 23.4$~mT at the graphene layer, located approximately 20~nm above the nanomagnet array.
These values represent a lower‑bound estimate of the stray field because (i) the NV measurements were performed at room temperature, whereas the ARPES experiments were conducted below 10~K, temperatures at which the $M_s$ of Co/Pt multilayers is expected to be higher than at room temperature~\cite{CoPt-MT}; and (ii) the assumed $M_s=300$~kA/m likely underestimates the true value, as it assumes that Pt remains nonmagnetic and thus dilutes the Co moments, despite the reality that a substantial induced magnetic moment in Pt layers can arise from proximity effects~\cite{CoPt-Prox-1,CoPt-Prox-2,CoPt-Prox-3,CoPt-Prox-4}.

% --- FIGURE 7 --- %
\begin{figure}
    \includegraphics[width=1\columnwidth,clip,angle =0]{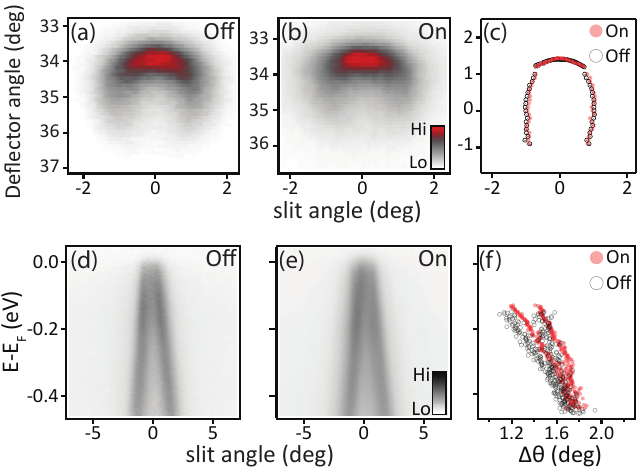}
    \caption{\label{fig7} 
    Comparison of ARPES spectra of graphene off [(a),(d)] and on [(b),(e)] the magnetic pattern. (a),(b) Constant energy contours around one Dirac cone. (c) Fitted trajectories from (a) and (b). (d),(e) Energy–momentum cuts at the graphene $K$ point. (f) Momentum broadening extracted from (d) and (e).
}
\end{figure}

Figure~\ref{fig7} compares ARPES spectra measured on and off the magnetic array.
Figures~\ref{fig7}~(a) and (b) show momentum-dependent spectral intensity at $E_B$~=~0.2~eV at one of the six Dirac cones.
Intensity on the side closer to $\Gamma$ point is expected to be higher due to the effects of the matrix element~\cite{ARPESmatrix}.
A constant angular offset of $0.4^\circ$ along the deflector direction is fully accounted for by a surface work-function difference between the two regions (see SI~\cite{SI}).
After correcting for this offset and aligning the Dirac point to $(0,0)$ [Fig.~\ref{fig7}~(c)] (see SI for the process to determine the position of the Dirac point~\cite{SI}), we find negligible emission angle contraction and no detectable rotation of the constant energy contour (CEC). By contrast, single dipolar magnet setups with $B_z \sim 20$~mT are known to induce more than 20\% angle contraction and a $\sim50^\circ$ CEC rotation under the same 37~eV photon energy, in addition to the detrimental momentum smearing due to a grossly enlarged virtual emission spot~\cite{B-ARPES-Rice,B-ARPES-ALS}.
%\\

To examine the momentum broadening effect, energy–momentum cuts at the $K$ points [Figs.~\ref{fig7}~(d) and (e)] show that the momentum distribution curve (MDC) widths $\Delta\theta$ differ by less than $0.2^\circ$ between magnetic and non-magnetic regions [Fig.~\ref{fig7}~(f)]. This broadening is negligible compared with the $2.9^\circ$ reported under $B_z = 3.2$~mT using earlier methods~\cite{B-ARPES-Rice}.

This result further demonstrates the advantage of using metamaterial-like nanomagnet arrays to \textit{in-situ} apply a magnetic field in the ARPES sample environment.
With the stray field attained in this demonstrative experiment approaching the condition for visualizing Landau quantization in momentum space (see SI for details~\cite{SI}), our approach underscores the potential to use this new approach to investigate field-induced electronic structure changes in a wide range of quantum materials. This includes potential momentum splitting in topological magnets with large effective orbital angular momentum, field-induced superconducting pair breaking, and superconducting vortex electronic structures.

% ----- 3 conclusion ------------------------------------------------- %

\section{Conclusion}
To conclude, we take advantage of recent advances in artificial spin ice design to realize an \textit{in-situ} sub-tesla magnetic field sample environment with nanomagnetic metamaterials.
The approach could provide a $\sim10$~mT to $\sim1$~T magnetic field in the ARPES sample environment while keeping the photoelectron trajectory nearly unaffected.
We demonstrate the practicality of implementing such a design in ARPES experiments and validate the predicted minimal photoelectron distortion effect.
Our work paves new ways to \textit{in-situ} implement versatile magnetic field patterns, and opens up possibilities for photoemission investigations of rich solid-state phenomena under strong magnetic fields.

% ----- A epilogue --------------------------------------------------- %
\begin{acknowledgments}
%NSF
The work at Yale University is partially supported by the National Science Foundation (NSF) under DMR-2239171.
% -- WL fellowship
W. L. acknowledges support from the James Kouvel Fellowship and the John F. Enders Fellowship.
% -- BC
The work at Boston College is supported by the Office of Naval Research (grant N00014-24-1-2102).
% -- ASI
The work at Princeton University and University of Minnesota is supported by the National Science Foundation (NSF) under DMR-2419407.
% -- Georgia
The work at Georgia Institute of Technology is supported by the U.S. Department of Energy (DOE), Office of Science, Basic Energy Sciences (BES), under award No. DE-SC0024870.
% -- SSRL
Use of the Stanford Synchrotron Radiation Lightsource, SLAC National Accelerator Laboratory, is supported by the U.S. Department of Energy, Office of Science, Office of Basic Energy Sciences under Contract No. DE-AC02-76SF00515.
% -- BC cleanroom
We also acknowledge that some of the work was carried out in the Boston College cleanroom and nanotechnology facilities.
\end{acknowledgments}

\bibliography{bibfile.bib}

% --- END ------------------------------------------------------------ %
\end{document}